\documentclass[preprint,showpacs,showkeys, prX]{revtex4-1}
\usepackage{hyperref}
\usepackage[utf8x]{inputenc}
\usepackage{amsmath}
\usepackage{amssymb}
\usepackage{mathptmx}
\usepackage{hyperref}
\usepackage{graphicx}
\usepackage{enumerate}
\usepackage[margin=2.5cm]{geometry}

\begin{document}
\title{Variance-based control of regime shifts: bistability and oscillations}
\author{\surname{Anselmo} Garc\'ia Cant\'u Ros, \surname{Till} Fluschnik, \surname{Juergen} Kropp}
\affiliation{Potsdam Institute for Climate Impact Research (PIK), Potsdam, Germany.}
\email{anselmo@pik-potsdam.de}
\begin{abstract}
A variety of real world and experimental systems can display a drastic regime shift, as the evolution in one its paramaters crosses a threshold value. Assimilation of
such a transition with a bifurcation has allowed to identify so called ``early warning signals'', at the level of the time series generated by
the system underscope. The literature in early warning detection methods is currently expanding and their potential for practical applicability is 
being discussed in different contexts. In this work, we elaborate on the use of the variance of a system variable, which constitutes the simplest early warning 
indicator, to gain control on the long-term dynamics of the system, while extending an exploitation phase. In particular, we address the cases of the 
cusp and Hopf normal forms, as prototypical examples of bistability and oscillations. Our results provide insights on the interplay between the time-scale for 
the system observation, the degree of sensitivity of the control feedback and the intensity of the random perturbations, in shaping the long-term control efficiency.
\end{abstract}

\maketitle

\section{Introduction}
Along the last few decades an increasing number of scholars has been contributing to develop a coherent and further 
integral body of knowledge on complex real world systems. Cross-fertilization of concepts and 
tools stemming from dynamical systems, statistical physics and information and computation theories has allowed to identify and analyse systems 
whose observable behaviors result from similar underlying mechanisms \cite{NicolisI,Politi}. As a result of these attempts, a key finding has been 
realization of the ubiquity of nonlinear feedbacks in the dynamics of different classes of systems. 

In particular, the occurence of drastic, qualitative shifts of the functioning
regime of systems --as resulting from strong positive and negative feedbacks-- has attracted interest in diverse research fields, 
as diverse as microscopic physical systems at equilibrium \cite{Stanley} and their macroscopic non-equilibrium counterpart \cite{NicolisII,Cross}, ecology \cite{SchefferIV,Scheffer}, socio-physics \cite{Castellano} and Earth system science \cite{Lenton,Scheffer}.

Insights into the phenomenology, related to regime shifts and the development of new methods for their early detection, have
 relied on minimal models. A frequent idea in this approach is to cast the description of the system evolution as a stochastic differential equation \cite{Gardiner}, of the form:
\begin{equation}
\textbf{dx}=\textbf{F}(\textbf{x},\lambda)dt+d\textbf{$\textbf{W}_x$}\label{EqGen1}
\end{equation}
The key variables of the system are represented as a vector $\textbf{x}=(x_1,x_2,\dots,x_N)$ and their dynamic interrelationships are considered as deterministic laws, 
described as a vector field \textbf{F}. Vector $\textbf{W}_x$ aims at introducing the effect of incessant perturbations on the \textbf{x} components, as typically resulting 
from endogenous meso-scale dynamic complexity and from an infinite number of uncorrelated 
environmental factors. Consequently, $\textbf{W}_x$ is assimilated to a Gaussian Wiener process \cite{Gardiner} with standard deviation $\omega_{x}$. The parameter $\lambda$ 
controls the qualitative behavior of solutions to the deterministic 
part of system \eqref{EqGen1} around a stationary state $\textbf{x}_s$. In this framework, a regime shift is represented as a local bifurcation occuring in the deterministic
part of the system at a critical point $(\textbf{x}_s,\lambda_c)$, i.e., that the Jacobian matrix of \textbf{F}, evaluated at $\textbf{x}_c$,  possesses one eigenvalue 
whose real part tends to zero as $\lambda$ approaches the value $\lambda_c$ \cite{Guckenheimer}. Thereby, the response of the system to random fluctuations can be studied 
along the evolution towards the bifurcation point.

Given the relevance of applications and implications of regime shifts in real world systems, it is of increasing interest to construct 
early warning indicators (\textbf{EWI}) \cite{SchefferI,SchefferII} -- with focus on identifying 
statistical signatures of the loss of stability in the time series generated by the evolution of the system towards criticality. Conjointly, 
a relevant phenomenon is the slowing down of the 
fast stabilizing dynamic modes --often referred as critical slowing down. This results from the real part 
of the leading eigenvalue becoming zero as the system closely approaches the bifurcation point \cite{SchefferI}. 

As illustrated in a pioneering work by S. Carpenter \cite{Carpenter} on \textbf{EWI}-based monitoring of shallow lake ecosystems, the onset of the critical slowing 
down can lead to a drastic increase in the variance of a variable of the system. More recently, the skewness \cite{Guttal}, kurtosis\cite{Biggs} and the autocorrelation function \cite{Livina} of a system variable have been shown to capture relevant information
on the increased loss of stability close to the bifurcation point.
  
The literature in early warning detection is rapidly expanding and their potential applications within a broad range of real world problems is being discussed \cite{SchefferI,SchefferII}.
Within this contextual frame, it becomes natural to inquire about the possible long-term complex behavior that would arise if \textbf{EWI}s are repeatedly used to control systems potentially undergoing regime shifts. 
As far as we know, this question
remains open, in the light of earlier and further recent advanced methods in the construction of \textbf{EWI}'s \cite{Dakos}.

In order to tackle this question, we propose to analyze the dynamics of a system represented
by \eqref{EqGen1} subject to a control process. The basic idea is that such a steering control should: 1) re-establish the stability of the system before it crosses a critical point, or 
2) re-establish a lost initial system regime once a critical point has been crossed, to then let the system evolve again towards criticality.

Let us explore this idea by considering a control system which reacts upon the behavior of an \textbf{EWI}, denoted by $I_{\tau}$, whose construction
 is based on information about the past states of the system within a sliding observation time 
window of size $\tau$. In other words, we are interested in analyzing the coupled dynamics of \eqref{EqGen1} with 
an evolution equation for $\lambda$, of the form
\begin{equation}
d\lambda=\Phi(I_{\tau},\alpha)dt+dW_{\lambda} \label{EqGen2}
\end{equation} 
Here we require the control functional $\Phi$ to operate a change of sign, at the level of $d\lambda$, 
once $I_{\tau}$ crosses a reference value $V_R$. In order to allow for differences in the degree of sensitivity of the control system,
let us consider the control response to changes in the ratio $I_{\tau}/V_R$ as modulated by a parameter $\alpha$. Moreover, let us assume that the control response
may exhibit random variability as a result of the influence of many uncorrelated factors. Thus, we denote by $W_{\lambda}$ a Gaussian Wiener process with 
standard deviation $\omega_{\lambda}$.

In the domain of application of \textbf{EWI}s, relevant cases are those where the control variable $\lambda$ is driven by human interventions. 
In many of these instances, the evolution of the system towards criticality is the result of an ``exploitation'' process, that is sought to be 
intensified and mantained, while avoiding a system regime shift. Consider for instance catching rates in sustainable fisheries \cite{Ray}, agriculture related leaching rates of phosphorus into a shallow 
lake, whose oligotrophic regime is to be preserved \cite{SchefferIII}, or the increase in 
the autocatalytic production of chemical species in a well-stirred reactor \cite{NicolisIII}, where a low entropy production regime may be desirable. 
Consequently, we characterize the dynamics of system \eqref{EqGen1}-\eqref{EqGen2} in terms of two phases: \textbf{\textit{exploitation}} if $\Phi>0$ and \textit{\textbf{recovery}} if $\Phi<0$.

Central questions for us are: 
\begin{itemize}
 \item[\textbf{\textit{i}})] The role of the time-scale $\tau$, considered for the construction of the \textbf{EWI}s, in 
  shaping the emergence of control properties in a system \eqref{EqGen1}-\eqref{EqGen2}. 
  \item[\textbf{\textit{ii}})] The long-term complex dynamics that can result
  from the temporal non-locality introduced by the \textbf{EWI} into the dynamics of the coupled system \eqref{EqGen1}-\eqref{EqGen2}. In other words,
  in the long-term, the control exerts changes on the system on the basis of the past dynamics of the system and its control as a whole. This opens the
  possibility to observe different complex behaviors, as the effective dimension of the system-control is augmented by its dependency on the past.
 \item[\textbf{\textit{iii}})] The efficiency of the control process in terms of: its capacity to increase and mantain explotation in a selected regime 
as a function of the observation window $\tau$, of the control response sensitivity $\alpha$ and of the variance of stochastic perturbations $\omega_x$ and $\omega_{\lambda}$.
\end{itemize}

In Sec.II we address these questions, by exploring numerically the performance of control on minimal dynamical systems of generic character, 
that reproduce the prototypical phenomena of bistability and oscillations. Each of these controlled systems is first studied in absence of noise,  
in order to reveal the role of the observation time scale in shaping the leading deterministic dynamics. Upon the basis of the insights thereby 
generated, for each system, we introduce a suitable measure of control efficiency. Subsequently, we address the influence of the variance of noise, 
for different observation time windows and different degrees of control response, on the efficiency of both systems under consideration. Finally, in Sec.III 
we summarize our main results and conclusions.

\begin{figure}
\begin{centering}
\includegraphics[width=.65\textwidth,height=.35\textheight]{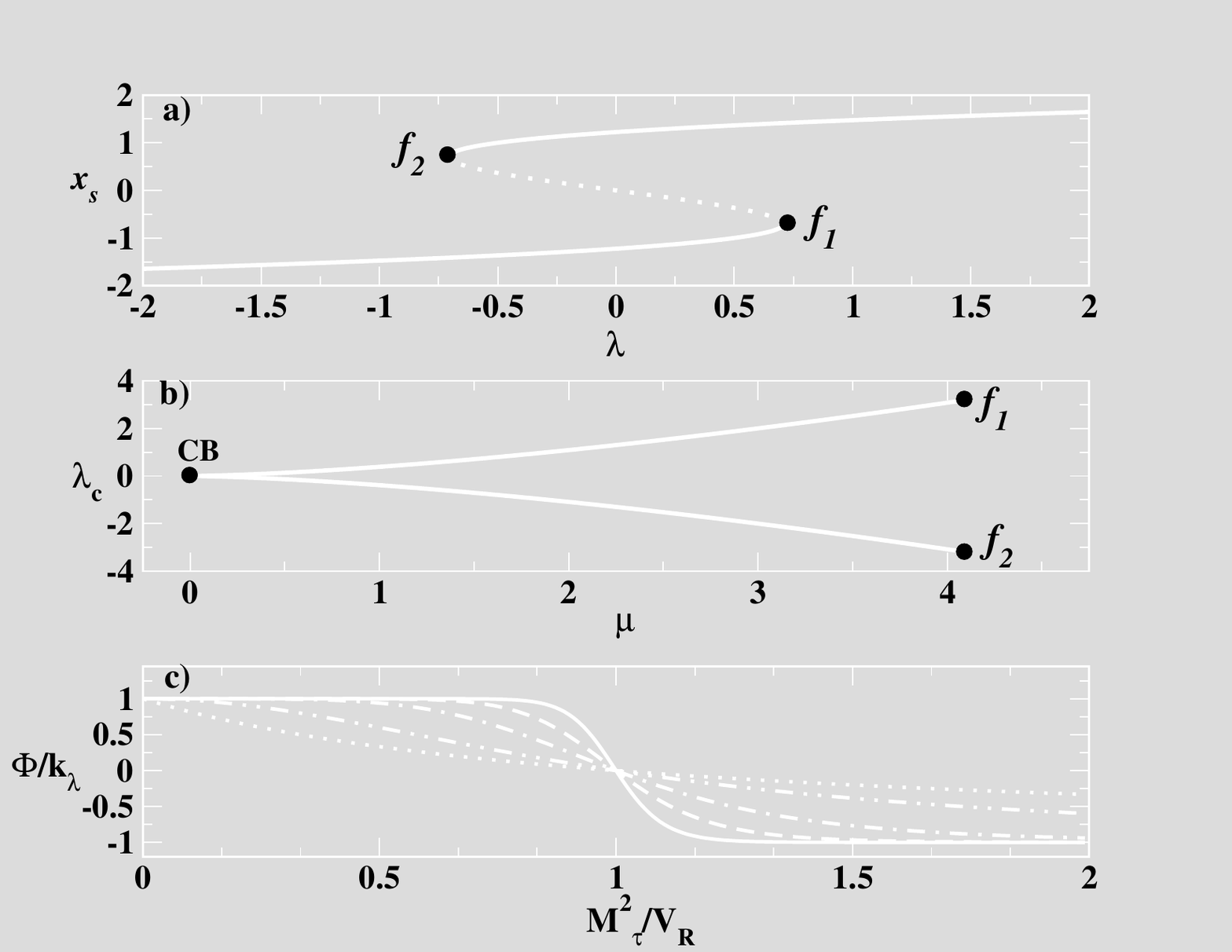}
\caption{ a) plot of a fold bifurcation diagram, as provided by the stationary solutions of the cusp normal form \ref{Cusp}, as a function 
of parameter $\lambda$ and for $\mu=3/2$. The solid lines correspond to the stable branches and the dashed line to the unstable one. Unstable and stable 
branches meet and anhihilate each other at the saddle-node bifurcation points $f_1$ and $f_2$; b) plot of the branches formed by the critical values of $\lambda$ corresponding
to both $f_1$ and $f_2$ bifurcation points, as a function of a second parameter $\mu$. Both branches converge to the cusp bifurcation point CB at $\mu=0$. Finally, 
plot $\Phi/k_{\lambda}$ vs $M^2_\tau/V_R$, for $\alpha=1$ (dot), $\alpha=2$ (dot-dot-dash), 
$\alpha=5$ (dot-dash), $\alpha=10$ (dash) and $\alpha=20$ (solid). }\label{Bif}
\par\end{centering}	
\end{figure}

\section{Normal forms and control}
Regardless of the complexity of its dynamics, as a system evolves further into the vicinity of a bifurcation point, there is a decrease in the number of degrees of freedom necessary
to effectively describe the dynamics. This remarkable fact can be understood as a direct consequence of the gradual run out of the fast-relaxation dynamic modes 
that occurs along with the critical slowing down. A corner stone in local bifurcation theory is the application of the centre manifold method \cite{Guckenheimer} to identifying
 canonical descriptions of the nonlinear behavior around a bifurcation point, in terms of a reduced number of variables that contain the whole information about the dynamics of the slow leading modes. 
Such simple representations --so called normal forms-- describe the long-term behavior of the normalized amplitude of the stable solutions that emerge at criticality. 
The importance of normal forms lies on the fact that the dynamics of different types of systems, around the critical point, can be mapped one-to-one continuously  
onto the one displayed by a normal form \cite{NicolisIV}. This property  --known as topological equivalence-- allows to define classes of universality, where near the critical point, the dynamics of systems belonging to the same class can be mapped onto each other by a 
change of variables preserving the direction of trajectories. Naturally, a simple and most representative element in a class of universality is thus the corresponding normal form.

With the aim of tackling questions \textbf{\textit{i}})-\textit{\textbf{iii}}) above in a general context, we shall focus our analysis of control 
dynamics on selected normal forms. As an \textbf{EWI} 
we consider the variance $M^2_{\tau}[\delta x_j]=\frac{1}{\tau}\int_{t-\tau}^{t} (\delta x_j(t'))^2 dt'$ of deviations of a system variable around its mean value, within the time window $\tau$, i.e. 
$\delta x_j=x_j-\frac{1}{\tau}\int_{t-\tau}^{t} x_j(t') dt'$. Consistently 
with the minimal character of the normal form description, let us consider a simple expression for the control functional in \eqref{EqGen2}, of the form
\begin{equation}
\Phi(\tau,\alpha)=k_{\lambda}\Bigg(1-2\frac{M_{\tau}^2[\delta x_j]^\alpha}{M_{\tau}^2[\delta x_j]^\alpha+V_R^\alpha}\Bigg) \mbox{, $k_{\lambda},\alpha>0$}\label{Control}
\end{equation}
 where $k_{\lambda}$ corresponds to the maximum exploitation and recovery rates. For low values in $\alpha$ both the exploitation and recovery phases develop weakly as $M_{\tau}^2[\delta x_j]$ crosses the threshold value $V_R$.
 As $\alpha$ is arbitrarily increased, the functional $\Phi$ sharply approaches a Heavyside function (see (Fig.\eqref{Bif}c)) --which emulates a full blown response of the control system \eqref{EqGen2} to a threshold overshooting
\begin{equation}
\Phi(\tau,\alpha\rightarrow\infty)=\left\{
\begin{array}{c l}
  k_{\lambda} & \mbox{$\frac{V_R}{M_{\tau}^2[\delta x_j]}>1$}\\ \\
 0 & \frac{V_R}{M_{\tau}^2[\delta x_j]}=1 \\ \\ -k_{\lambda} & \mbox{otherwise} \label{ControlI}
\end{array}\right. 
\end{equation} 
Here, a comment is in order with regard to the choice of the variance of a system variable as an \textbf{EWI}. As discussed in \cite{DakosI}, in the more general case where the standard deviation of the random perturbations is 
a time dependent function, the variance of a system variable may not 
constitute a reliable \textbf{EWI}. Consequently, we shall restrain ourselves to the case where both standard deviations $\omega_x$ and $\omega_{\lambda}$ in \eqref{EqGen1}-\eqref{EqGen2} are constants.

\begin{figure}
\begin{centering}
 \includegraphics[width=.65\textwidth,height=.35\textheight]{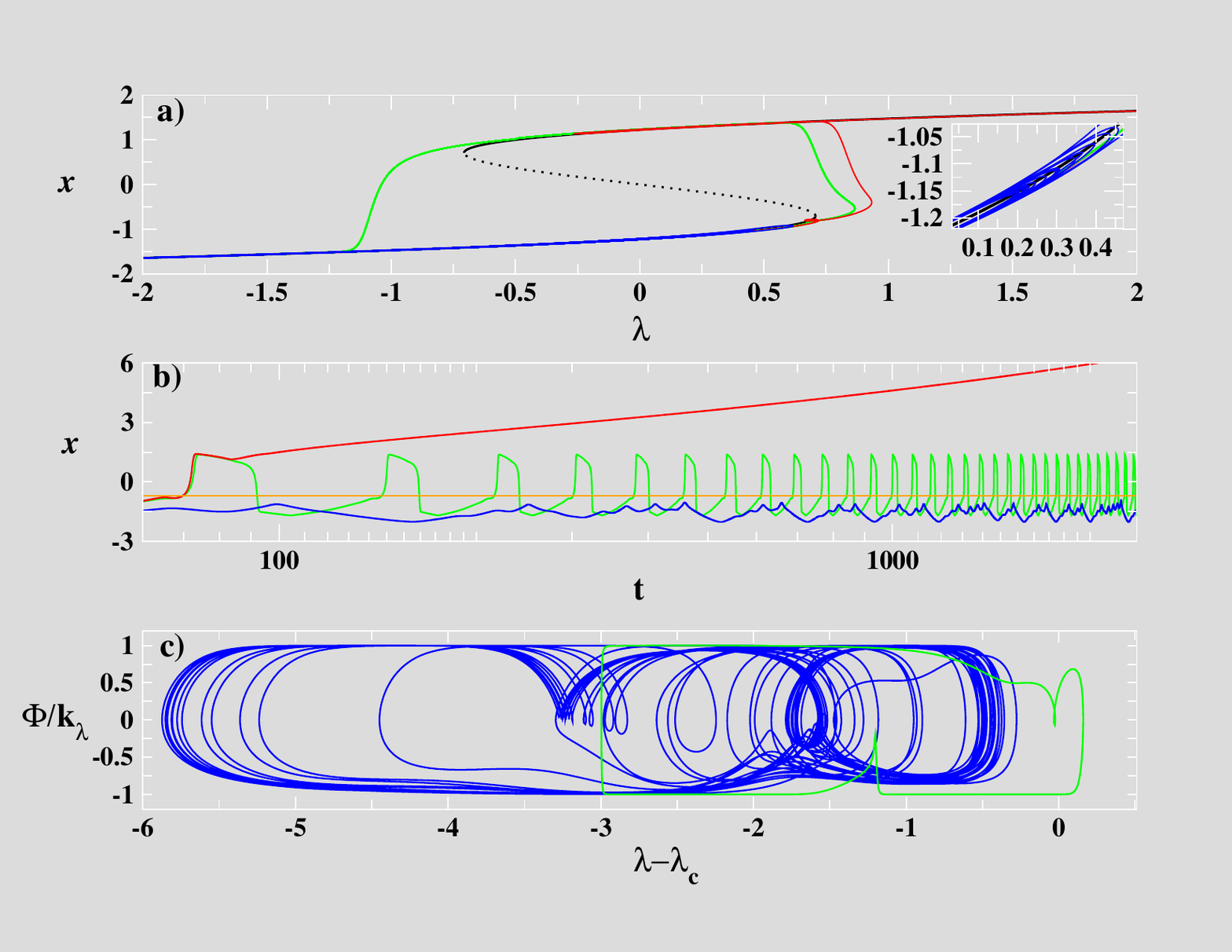}
\caption{For $\tau=10$ (red), $\tau=11$ (green) and $\tau=35$ (blue), a) plot of a \textbf{CCS} deterministic trajectories ($\omega_x=0, \omega_{\lambda}=0$) 
in the ($\lambda,x$)-plane (in black lines the bifurcation diagram of the cusp normal form); b) plot \textit{x} \textit{vs} time (in orange the critical value $x_c$) and c) Plot of $\Phi/k_{\lambda}$ \textit{vs} $\lambda-\lambda_c$, as related to the system trajectories in panels a) and b),
for $\tau=11$ and $\tau=35$. Here, values $\Phi>0$ and $\Phi<0$ are considered as exploitation and recovery, respectively. Parameters and initial condition: $\mu=3/2$, $k_{\lambda}=1/10$, $\alpha=2$, $V_R=1/100$. $x(0)=-3/2$, $\lambda(0)=-5$. Time step $dt=5\times 10^{-3}$, numerical integration over $5\times 10^{5}$ time steps.}\label{Traj}
\par\end{centering}	
\end{figure}

\subsection{The cusp-control system (CCS)}
As pointed out in \cite{SchefferI,Hastings}, the phenomenology observed in so called catasrophic regime shifts in different types of systems can be suitably assimilated 
with a saddle-node bifurcation. As a parameter evolves, a saddle-node bifurcation occurs when a stable (node) and an unstable (saddle) fixed point solution, to the deterministic evolution 
law of the system, approach each other and anhihilate upon meeting. This entails the destabilization
of the system, which is thus led to an abrupt runaway --this constitutes a catastrophic regime shift. However, since in real systems unbounded runaways are 
ruled out, saddle-node bifurcations are typically found in cases where the branch formed by the system fixed points folds upon itself, within a range of 
values in the control parameter $\lambda$ (see Fig.\ref{Bif}a). The presence of such a folding entails the existence of a region where two stable solution branches coalesce along with an unstable one. In this picture, 
the unstable branch marks the 
boundary line between the two basins of attraction, associated to both stable branches. In general, the distance between the bifurcation points at the extremes of the s-shape folding ($f_1$ and $f_2$ in Fig.\ref{Bif}b) 
is determined by a 
second paramater $\mu$. Examples of such a bistability are reported in different types of systems, ranging from laser and cell division, to ecoystems and climate \cite{SchefferI}. 

In order to explore the question of bifurcation control in the context of bistability, let us consider the cusp normal form
\begin{equation}
F(x,\lambda)=x(\mu \pm x^2) +\lambda \label{Cusp} 
\end{equation}

The solutions to the stationary form $F(x_s,\lambda)=0$ (with the sign minus in \eqref{Cusp}), as a function of $\lambda$, correspond to those plotted in Fig.\ref{Bif}a. The
position of the saddle-node bifurcation points $f_{1,2}$ are determined by the relation $f_{1,2}=\{(\lambda,\mu)|\lambda=\pm\frac{2}{3\sqrt{3}}\mu^{3/2}\}$. 
Accordingly, both branches approach each other and merge as $\mu$ tends to zero, at the so-called cusp bifurcation point (Fig.\ref{Bif}b). A similar behavior occurs when changing the sign in the second term in \eqref{Cusp}, which amounts
to inverting the s-shaped folding. 
  
We carry out the numerical integration of the variance-based cusp-control system
(\textbf{CCS}) (\ref{EqGen1}), (\ref{EqGen2}), (\ref{Control} and \ref{Cusp}) via a standard Euler-Maruyama approximation \cite{Gardiner}.

\subsubsection{CCS deterministic dynamics}
In order to illustrate the dynamics of the \textbf{CCS} in absence of noise ($\omega_x=0, \omega_{\lambda}=0$), let us consider exploitation 
as initially occuring along the negative branch in Fig.\ref{Bif}a, a moderate degree of control response $(\alpha=5)$ (see Fig.\ref{Bif}c) and a small variance reference threshold
$V_R=1/100$.

 According to the numerical results, we distinguish three basic long-term \textbf{CCS} behaviors:
\begin{itemize}
 \item[\textbf{(B1})] \textit{Exploitation runaway}: 
 If the time window is small as compared to a critical value 
$\tau_c$, the system crosses the bifurcation point $f_1$ and a regime shift occurs. As the system is expelled onto the upper branch,
 the control mechanism reacts to the overshooting of the reference value $V_R$. This triggers a recovery stage ($\Phi<0$) along the upper branch. However, as the system evolves along the upper branch,
 it eventually losses completely the ``memory'' on the original low branch regime. As a result, the variance decreases gradually below $V_R$, 
 recovering is interrupted and the \textbf{CCS} switches to runaway explotation (red lines in Figs.\ref{Traj}a,b). 
 \item[\textbf{(B2})] \textit{Hysteretic pathways}: 
When the observation window is now greater than the critical value $\tau_c$, a transition occurs, from runaway exploitation to long-term control. For specific 
$\tau$-ranges, the system crosses the bifurcation point $f_1$ and it starts recovering along the upper branch. However, differently from \textbf{B1}, the upper branch 
recovering continues further and the bifurcation point $f_2$ is crossed. Thus, the system undergoes again a regime shift towards the original lower branch. As the system stability increases and information about
the upper branch is lost, the \textbf{CCS} eventually switches to exploitation, beyond $f_1$ (green lines in Figs.\ref{Traj}a,b). 
This exploitation-recovery process is mantained in the long-term. Processes describing cycles along the zone of bistability are commonly refered as
hysteretic.
Figure \ref{Traj}c depicts the long-term hysteretic exploitation-recovery process, as a function of the proximity to the critical value $\lambda_c$ corresponding
to the bifurcation point $f_1$. 
 \item[\textbf{(B3})] \textit{Branch-confined pathways}: For large $\tau$ values, escapes to the upper branch are 
supressed and the system remains oscillating aperiodically without crossing the bifurcation point $f_1$ (blue line, Figs.\ref{Traj}a,b). This aperiodic dynamics is predictable 
in the sense that the average growth of an initially small deviation around a reference state \cite{NicolisI} is sub-exponential (not shown). The increase in complexity of the dynamics 
occuring along with $\tau$ is 
illustrated by the inset of Fig.\ref{Traj}a and by Fig.\ref{Traj}c (blue lines). The overall range of exploitation and recovery becomes larger as $\tau$ 
is increased Fig.\ref{Traj}c. In contrast with case \textbf{B1}, the evolution towards
the bifurcation point is marked by short exploitation-recovery-exploitation cycles Fig.\ref{Traj}c, where effective recovery avoids a system regime shift. For certain $\tau$-ranges 
long-term intermittence between behaviors \textbf{B2} and \textbf{B3} arises (not shown).  

\end{itemize}

It is worth noticing from Fig.\ref{Traj}a that the regime shifts 
occur past and not at the crossing of the critical point $(x_c,\lambda_c)$ --following apparent
extensions of the lower stable branch, beyond the critical point. Such an ``inertia'' effect is a result of the critical slowing down, where the
dynamics of \textit{x} becomes 'slaved' by the slower dynamics of $\lambda$.

\begin{figure}
\begin{centering}
 \includegraphics[width=.65\textwidth,height=.35\textheight]{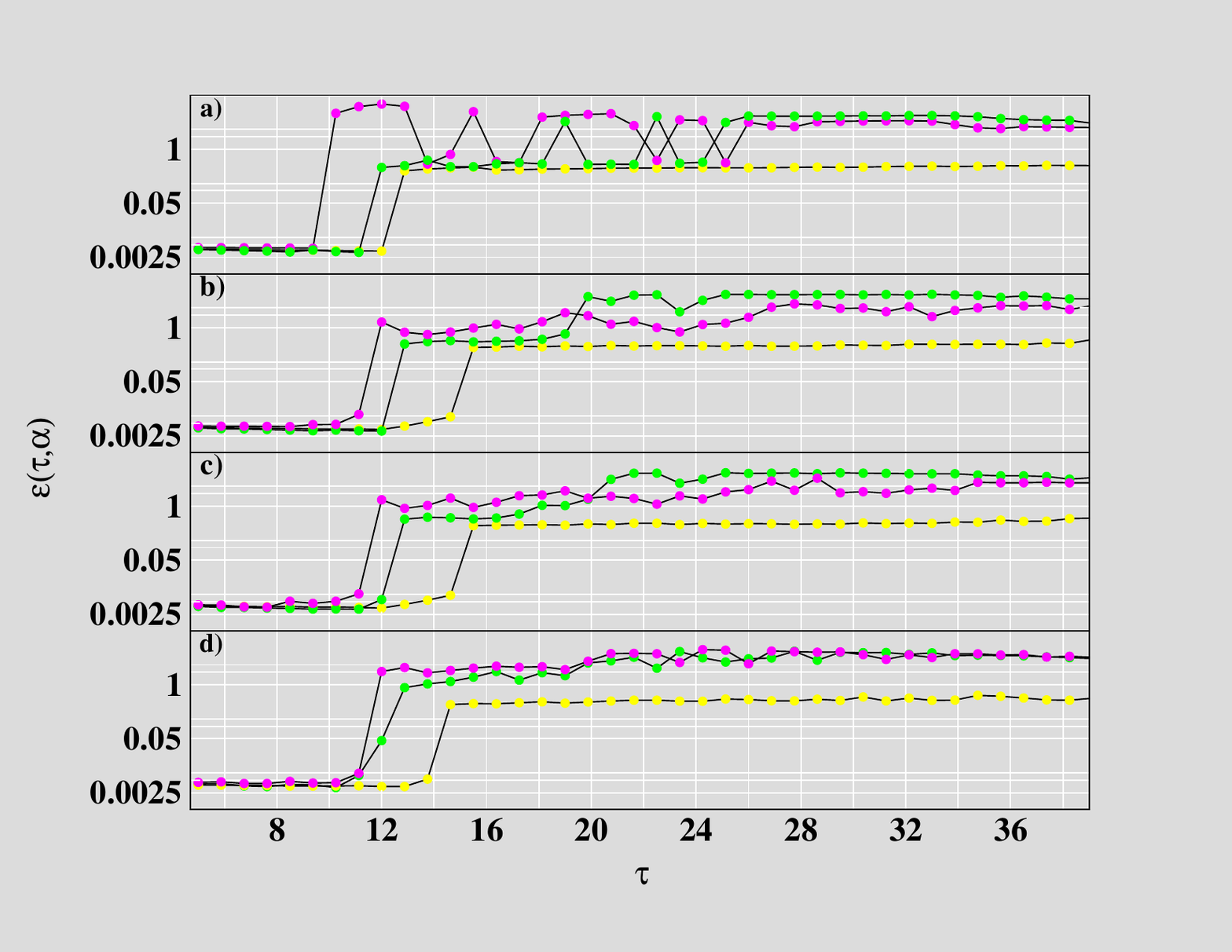}
\caption{For $\alpha=1$ (yellow), $\alpha=5$ (green) and $\alpha=20$ (magenta) and $\omega_{\lambda}=0$, a) plot of $\epsilon$ vs $\tau$
 for deterministic \textbf{CCS} trajectories ($\omega_x=0$). Similar plots for stochastic trajectories with b) $\omega_x=5$,  c) $\omega_x=10$ and d) $\omega_x=25$. Parameter values, initial condition and numerical integration as in Fig.\ref{Traj}. Here, each 
$\epsilon$ value has been computed for $1\times10^6$ time steps.}\label{Eff}
\par\end{centering}	
\end{figure}

\subsubsection{CCS control efficiency}
A meaningful way to characterize further the \textbf{CCS} dynamics is to quantify the control capacity to increase and mantain long-term explotation within a selected stable branch, 
as a function of the model paramaters. Accordingly, we quantify the efficiency as
\begin{equation}
 \epsilon(\tau,\alpha)=\lim_{T\rightarrow \infty} \frac{\omega^2_R}{T M_{T}^2} \int_{0}^{T} \frac{\Phi(\tau,\alpha) H(\Phi(\tau,\alpha)) H(-x(t))}{k_\lambda} dt\label{Effi}
\end{equation}
\textit{H} denotes a Heavyside function, whose value is zero for negative argument 
or one otherwise. The product of the Heavyside functions in the integral plays the role of an AND boolean operator, entailing that the contributions to 
the time average of $\Phi$ are considered only if $\Phi$ is positive and $x$ negative. The variance of deviations from the initial to the final time \textit{T}, in the denominator of \eqref{Effi}, aims at amplifying differences between
hysteretic (\textbf{B2}) and branch-confined (\textbf{B3}) control regimes. Without loss of generality, hereafter we set the reference value $\omega^2_R=1$ in \eqref{Effi}.

 In the next subsections we focus on the behavior of the efficiency as a function of the size of the observation window $\tau$, of the degree of control response
 $\alpha$ and of the variance of random fluctuations $\omega_x$ and $\omega_{\lambda}$. 

\subsubsection{Deterministic CCS efficiency}
\begin{figure}
\begin{centering}
 \includegraphics[width=.65\textwidth,height=.35\textheight]{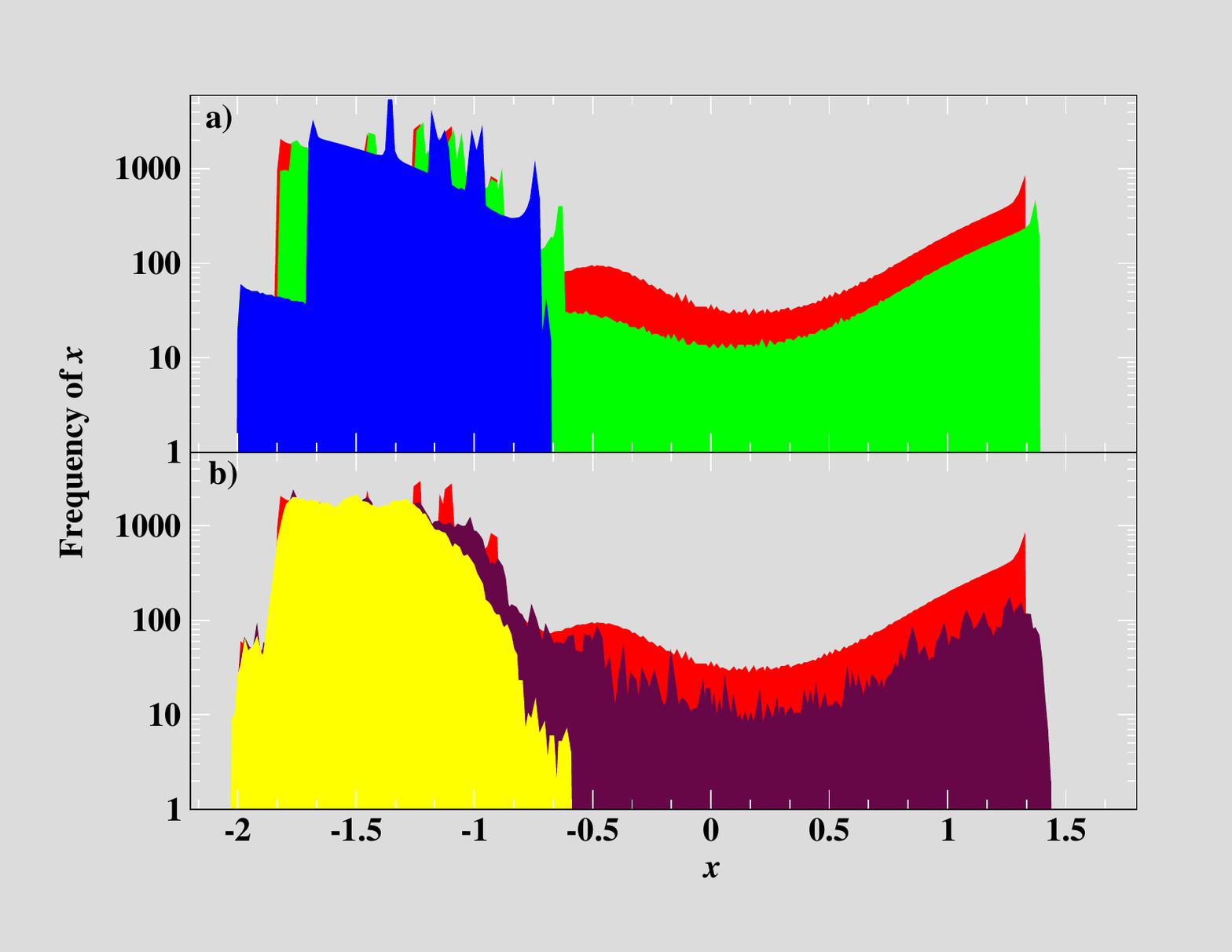}
\caption{ For $\alpha=20$, a) frequency histograms of \textbf{CCS} deterministic trajectories ($\omega_x=0$, $\omega_\lambda=0$) for $\tau=22.5$ (red), $\tau=22$ (green) and $\tau=19.5$ (blue). For 
$\tau=22.5$ and $\omega_\lambda=0$, b) histograms of stochastic trajectories with $\omega_x=10$ (brown) and $\omega_x=25$ (yellow), as well as
histogram of the deterministic case (red), as
in panel a). Parameter values, initial condition and numerical integration as in Fig.\eqref{Traj}.} \label{EffHist}
\par\end{centering}	
\end{figure}

Figure \ref{Eff}a summarizes the behavior of the efficiency \eqref{Effi} over the deterministic \textbf{CCS} trajectories corresponding to different values in $\tau$ and $\alpha$. 
Efficient control rises sharply as the observation window is increased above an $\alpha$-specific threshold $\tau_c$ --since below this value 
the \textbf{CCS} undergoes exploitation runaway (\textbf{B1}). Above $\tau_c$, the behavior of the efficiency becomes highly dependent on the degree of control response. 
For a low control response $(\alpha=1)$, the efficiency exhibits a quasi-constant plateau for $\tau>\tau_c$. This plateau consists of 
hysteretic cases (\textbf{B2}). In cases of moderate ($\alpha=5$) and high ($\alpha=20$) degree of control response, the behavior of the \textbf{CCS} efficiency becomes non-trivial. 
For values $\tau\gg\tau_c$ the efficiency is remarkably enhanced (see the appearance of a 
second plateau occurring for $\tau\geq 25$ in Fig\ref{Eff}a), as a result of the suppression of the \textbf{CCS} hysteretic control pathways, in favour of branch-confined recovery-exploitation cycles (\textbf{B3}). 
For intermediate values $\tau>\tau_c$, a combination of pathways 
 (\textbf{B2}) and (\textbf{B3}) is observed --where low efficiency values correspond to 
hysteretic control pathways (\textbf{B2}). Such a $\tau$-dependent selection of control pathways is further illustrated by the histogram 
of the \textit{x} variable in Fig.\ref{EffHist}a, for the high control response case. It shows that the dynamics of the \textbf{CCS} becomes 
confined to $\textit{x}<0$ values as $\tau$ is slightly shifted, which explains the alternace of low and high efficiency ranges appearing at 
intermediate $\tau$ values. Comparing the different control response cases in Fig.\ref{Eff}a, it can be 
observed that the moderate one is the most efficient for $\tau \gg \tau_c$. 
 
\subsubsection{Stochastic CCS efficiency}

Let us address now the case where the evolution of \textit{x} is marked by continuous random perturbations (i.e., $\omega_x\neq0$, $\omega_{\lambda}=0$ in Eqs. (\ref{EqGen1}) and (\ref{EqGen2})). 
Panels b) to d) in Fig.\eqref{Eff} illustrate the influence of noise on the \textbf{CCS} efficiency. In the case of low control response, the behavior of the efficiency with $\tau$ 
appears to be independent of the standard deviation of noise $\omega_x$. In contrast, for moderate and high control responses, the overall efficiency becomes enhanced by an increase in the
intensity of noise, 
in the sense that low efficiency values tend to disappear while high efficiency occurs all the way above the $\tau_c$-value (compare Fig.\ref{Eff}a with Figs. \ref{Eff}b--d). However,
if the variance of noise is marginally large (Fig.\ref{Eff}a-c), the high control response case is less efficient as 
compared with the moderate one. For strong noise (Fig.\ref{Eff}d), both moderate and high control response cases converge to a similar efficiency value for large $\tau$.

Since, for high and moderate response cases, low efficiency values tend to disappear when increasing intensity noise, it 
is natural to inquire about the influence of strong noise on hystheretic pathways. This effect is shown by the histogram of x Fig.\ref{EffHist}b for a single 
realization with $\alpha=20$. Notice the similarity between panels (a) and (b), which illustrates the equivalence in the control pathway selection that 
occurs either by changes in $\tau$ (Fig.\ref{EffHist}a) or by an increase in 
the standard deviation of the random fluctuations $\omega_x$ (Fig.\ref{EffHist}b).

\begin{figure}
\begin{centering}
 \includegraphics[width=.65\textwidth,height=.35\textheight]{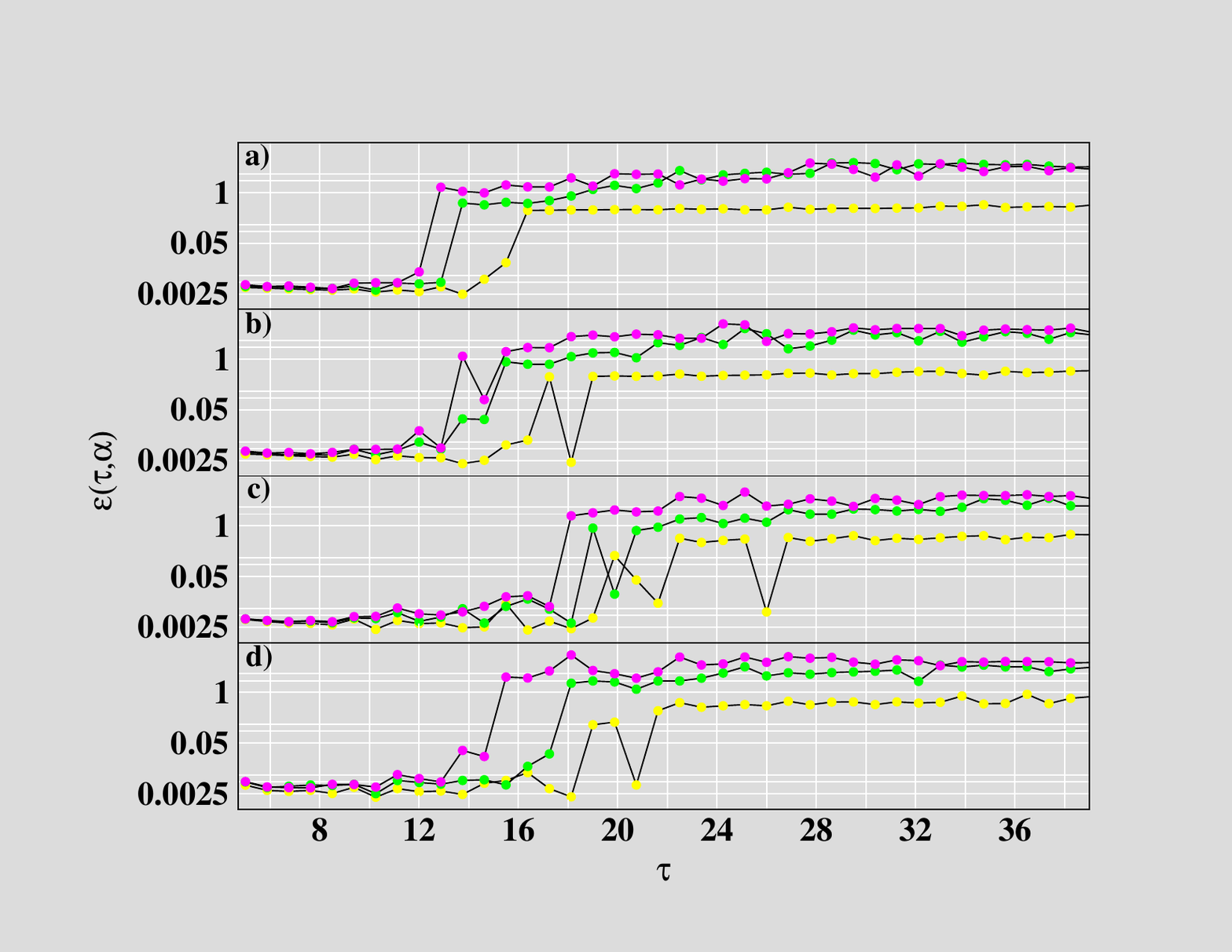}
\caption{For $\alpha=1$ (yellow), $\alpha=5$ (green) $\alpha=20$ (magenta) and $\omega_x=0$, plot of $\epsilon$ vs $\tau$
 for stochastic \textbf{CCS} trajectories with a) $\omega_{\lambda}=5$, b) $\omega_{\lambda}=10$ and c) $\omega_{\lambda}=20$. In d), similar plots for $\omega_{\lambda}=25$ 
and $\omega_x=20$. Parameter values, initial condition and numerical integration as in Fig.\ref{Traj}. Here, each 
$\epsilon$ value has been computed for $1\times10^6$ time steps.} \label{EffLambda}
\par\end{centering}	
\end{figure}

Interestingly enough, we observe a similar effect of efficiency enhancement and uniformization in the case of noise at the level of the control system 
(i.e., $\omega_x=0$, $\omega_{\lambda}\neq0$) Fig.\ref{EffLambda}a-c, as
compared with the deterministic \textbf{CCS} (Fig.\ref{Eff}a). However, in this case the emergence of efficient control occurs for larger values in $\tau$, 
as the intensity of noise increases. Finally, in the case of strong noise at the level of both the state variable \textit{x} and control (i.e., $\omega_x\gg0$, $\omega_{\lambda}\gg0$) Fig.\ref{EffLambda}d, 
the case of high control response becomes the most efficient one, in comparison with the ($\omega_x\neq0$, $\omega_{\lambda}=0$) cases (Fig.\ref{Eff}d).

\begin{figure}
\begin{centering}
 \includegraphics[width=.65\textwidth,height=.35\textheight]{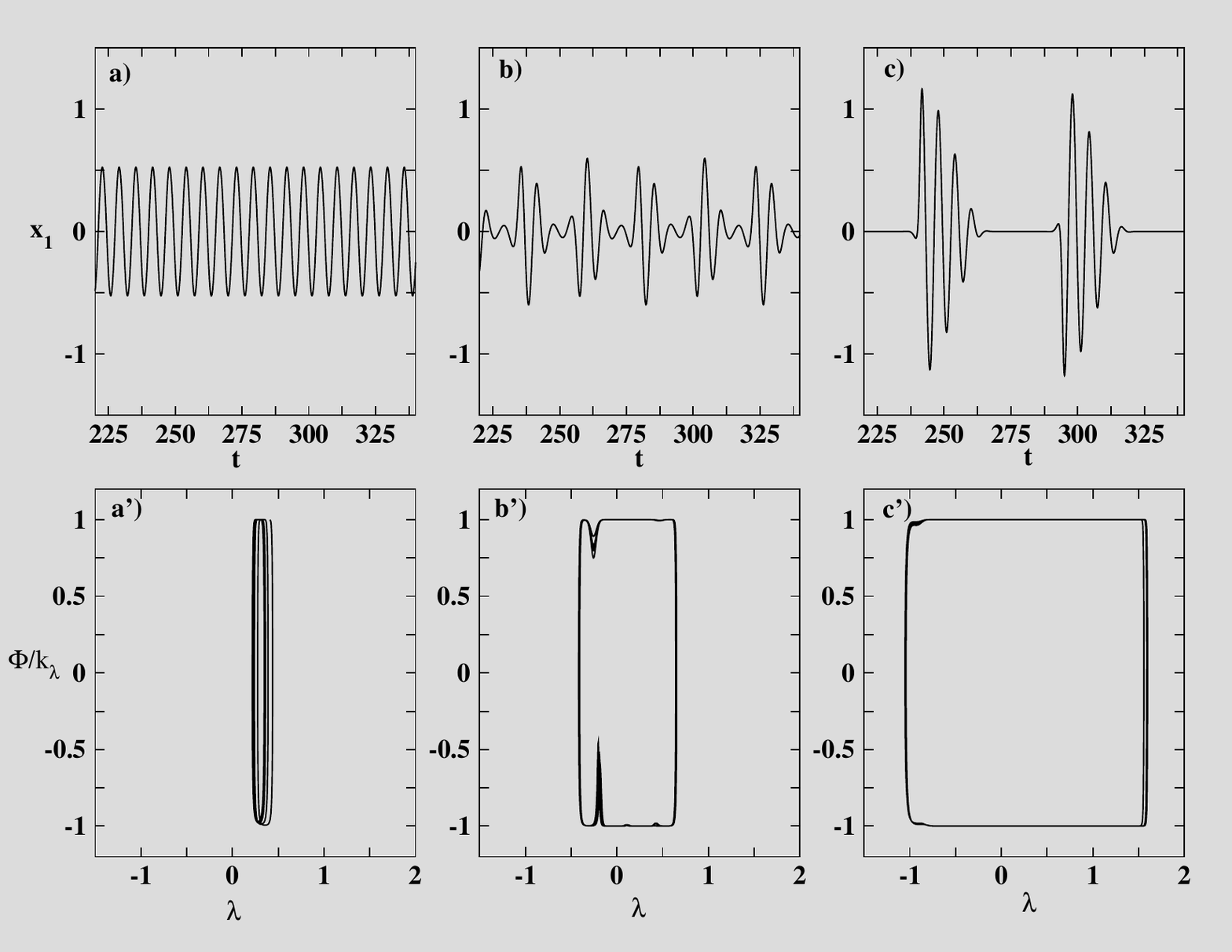}
\caption{ Behavior of the deterministic \textbf{HCS} ($\omega_x=0, \omega_{\lambda}=0$). Plot of $x_1$ \textit{vs} time for a) $\tau=0.5$, b) $\tau=3$, c) $\tau=10$. 
Corresponding to these $\tau$ values, panels a')-c') show the plot $\Phi/k_{\lambda}$ \textit{vs} $\lambda$.} \label{OscilaPhase}
\par\end{centering}	
\end{figure}

\subsection{The Hopf-control system (HCS)}
A relevant and also frequently observed type of regime shift consists in the onset of time periodic behavior, 
as a stationary system regime becomes unstable at a threshold value in a system parameter. Examples of processes and systems where such phenomenon 
is present are for instance, chemical clocks at the mesoscopic \cite{McEwen} and macroscopic \cite{Epstein} scales, cellular biochemical cycles \cite{Goldbeter}, prey-predator population 
dynamics \cite{Murray}, shallow lake ecosystems \cite{Scheffer}, the ocean-atmosphere system \cite{Vallis}, or non-equilibrium economics models \cite{Hallegatte}. All models 
accounting for the emergence 
of oscillatory behavior involve at least two variables and in all cases it has been possible to assimilate the mechanisms underlying the critical onset of oscillations 
to a Hopf bifurcation \cite{Guckenheimer}. In what follows we focus on the dynamics of the coupled system of control \eqref{EqGen2}-\eqref{Control}
and the deterministic evolution law in \eqref{EqGen1}, given by the Hopf normal form --which we refer hereafter as the Hopf-control system (\textbf{HCS}). 
For illustrative purposes and without loss of generality, we consider the supercritical version of the Hopf normal form
\begin{equation}
\begin{array}{c l}
F_1(x_1,x_2,\lambda)=\lambda x_1-x_2-x_1(x_1^2+x_2^2), & F_2(x_1,x_2,\lambda)=\lambda x_2-x_1-x_2(x_1^2+x_2^2) 
\end{array}\label{Hnormalform}
\end{equation}
The stationary solution $x_1=0$, $x_2=0$ to the Hopf normal form \eqref{Hnormalform} 
losses stability as $\lambda$ approaches zero along the negative axis. Above the critical value $\lambda_c=0$, the system variables $(x_1,x_2)$ describe stable oscillations
whose amplitudes increase as $\sqrt{\lambda}$. 

Here, the control functional \eqref{Control} is fed by either of both system variables $(x_1,x_2)$, whose dynamics is subjected to 
stochastic perturbations with same standard deviation $\omega_x$.

\subsubsection{Deterministic HCS dynamics}
For an initial condition around the stationary solution $x_1=0$, $x_2=0$ and for an intial value $\lambda<0$, Figs.\ref{OscilaPhase}a-c show the deterministic 
dynamics of the $x_1$ component under the influence of control for different values of the time window $\tau$. In complete absence of control, the linear growth in the value of 
$\lambda$, beyond the 
critical point, leads to oscillations of increasing amplitude $\sqrt{\lambda}$. In contrast, for small values $\tau\gtrsim0$ the amplitude of 
oscillations becomes constant (Fig.\ref{OscilaPhase}a). In this situation, both recovery ($\Phi<0$) and 
exploitation ($\Phi>0$) phases succeed each other rapidly, while the control variable always remains above the critical value $\lambda_c=0$ (Fig. \ref{OscilaPhase}a').
A further increase in $\tau$ induces a temporal alternance of small and large oscillation amplitudes (Fig.\ref{OscilaPhase}b), as 
the recovery and exploitation phases lead the system back and forth accross $\lambda_c$ (Fig.\ref{OscilaPhase}b'). If $\tau$ is large enough, oscillations are 
temporally suppressed (Fig.\ref{OscilaPhase}c), during a time interval that increases with $\tau$. This situation results from the fact that the $\lambda$-interval covered by the 
recovery and exploitation phases becomes enlarged (Fig.\ref{OscilaPhase}c'). In contrast with the \textbf{CCS} Fig.\ref{Cusp}c, no substantial
increase in the complexity of the dynamics is observed as the time window $\tau$ is arbitrarily increased.

\begin{figure}
\begin{centering}
 \includegraphics[width=.65\textwidth,height=.35\textheight]{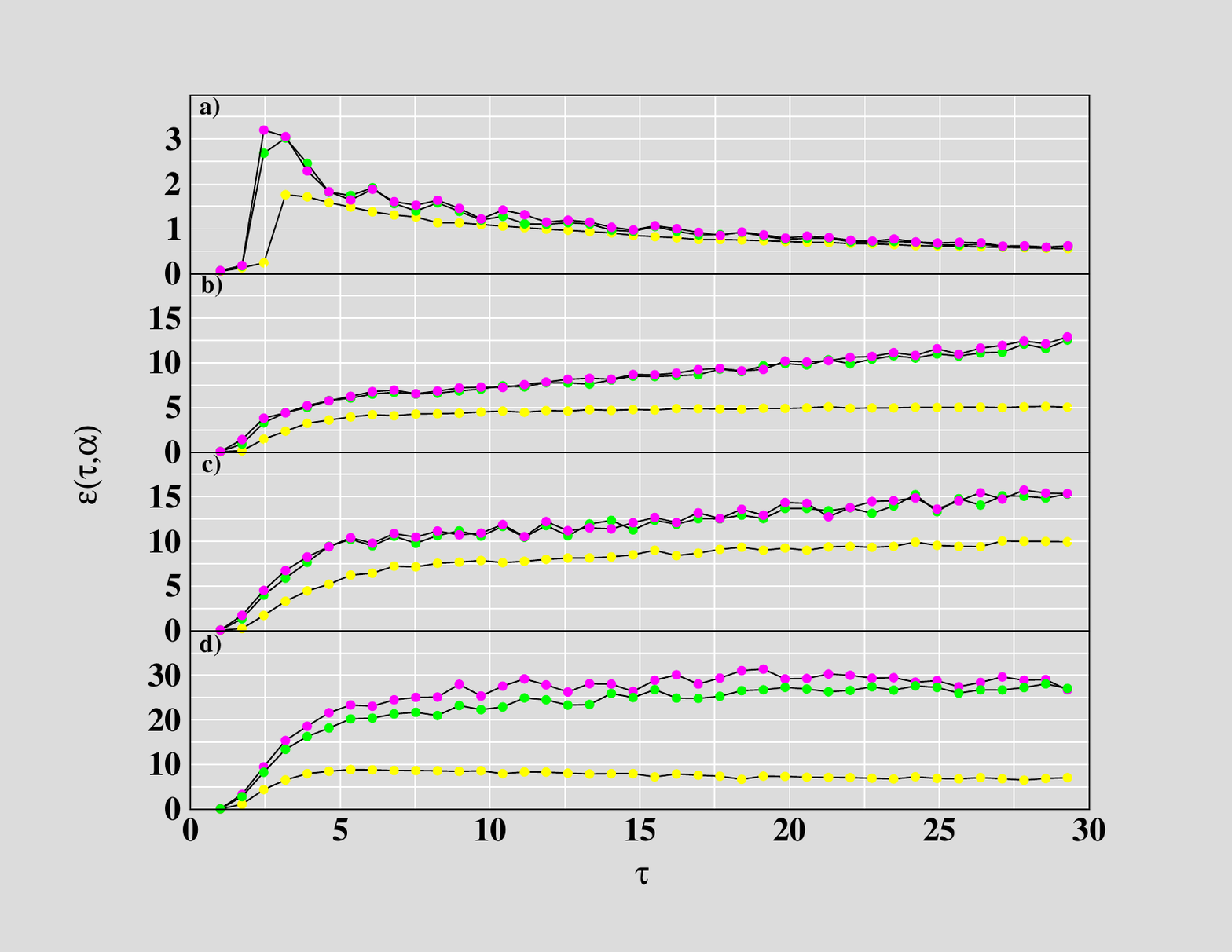}
\caption{For $\alpha=1$ (yellow), $\alpha=5$ (green), $\alpha=20$ (magenta) and $\omega_{\lambda}=0$, a) plot of $\epsilon$ vs $\tau$
 for deterministic \textbf{HCS} trajectories ($\omega_x=0$). Similar plots for stochastic trajectories with b) $\omega_x=5$,  c) $\omega_x=10$ and d) $\omega_x=25$. Parameter values, initial condition and numerical integration as in Fig.\ref{Traj}. Here, each 
$\epsilon$ value has been computed for $1\times10^6$ time steps.}\label{EffHopf}
\par\end{centering}	
\end{figure}

\subsubsection{Deterministic and stochastic HCS efficiency}
In a similar vein as for the \textbf{CCS}, we define the efficiency of the \textbf{HCS} as the capacity to increase and mantain long-term 
subcritical explotation, while keeping the oscillation amplitudes to a minimum. Thus we write the efficiency as   
\begin{equation}
 \epsilon(\tau,\alpha)=\lim_{T\rightarrow \infty} \frac{\omega^2_R}{T M_{T}^2} \int_{0}^{T} \frac{\Phi(\tau,\alpha) H(\Phi(\tau,\alpha))H(-\lambda(t))}{k_\lambda} dt\label{EffiH}
\end{equation} 
where H denotes a Heavyside function, such as defined for \eqref{Effi}.
\\
As shown by Figs.\ref{OscilaPhase}a'--c', subcritical exploitation increases with the size of the observation window $\tau$. However, it is also observed that
the amplitude of oscillations along the recovery phase also increases with $\tau$ (Figs.\ref{OscilaPhase}a--c). Therefore, an optimum is expected to occur in the efficiency \eqref{EffiH} as 
a function of $\tau$. This situation is actually illustrated in Fig.\ref{EffHopf}a for different degrees of control response. Notice that compared to the efficiency of the
deterministic \textbf{CCS} Fig.\ref{Effi}a, the sharp rise towards efficient control occurs at a small value $\tau_c$, above which efficiency gradually decays for larger
$\tau$ values. 

\begin{figure}
\begin{centering}
 \includegraphics[width=.65\textwidth,height=.35\textheight]{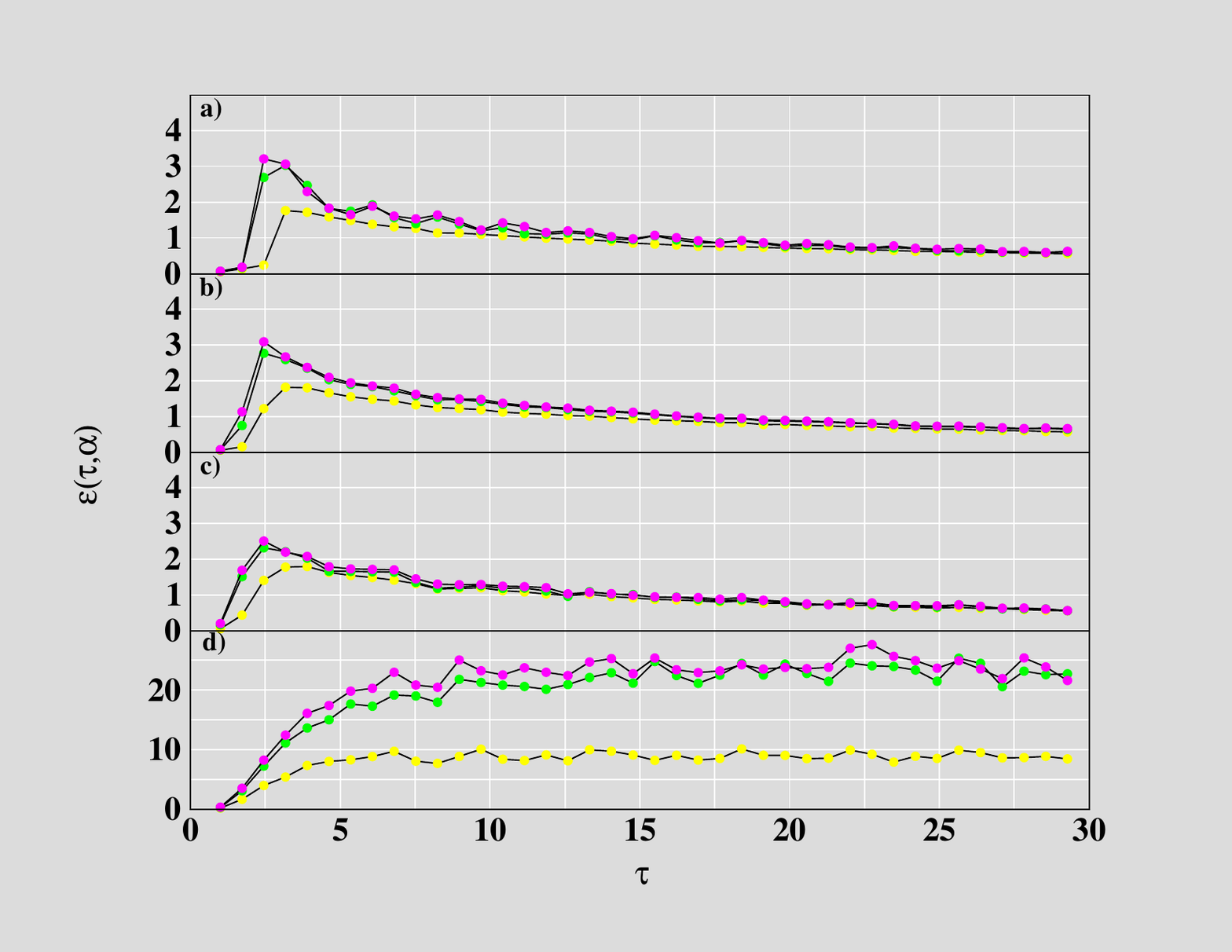}
\caption{For $\alpha=1$ (yellow), $\alpha=5$ (green) $\alpha=20$ (magenta) and ($\omega_x=0$), plot of $\epsilon$ vs $\tau$
 for stochastic HCS trajectories with a) $\omega_{\lambda}=5$, b) $\omega_{\lambda}=10$ and c) $\omega_{\lambda}=20$. In d), similar plots for $\omega_{\lambda}=25$ 
and $\omega_x=20$. Parameter values, initial condition and numerical integration as in Fig.\ref{Traj}.} \label{EffHLambda}
\par\end{centering}	
\end{figure}

 In presence of noise at the level of the system variables ($x_1,x_2$) (i.e. $\omega_x\neq0, \omega_{\lambda}=0$), it is observed that the efficiency 
increases along with $\tau$, for the moderate and high control response cases, while for low control response it remains at low practically constant value (Fig.\ref{EffHopf}b). 
Similarly as for the \textbf{CCS} (Fig.\ref{Eff}), the efficiency is in overall enhanced by the intensity of noise Fig.\ref{EffHopf}b--d. 
Moreover, the behavior of the efficiency is practically the same for moderate and high control response cases, with the exception of the strong noise case  Fig.\ref{EffHopf}d,
 where high control response is highest for arbitrary large $\tau$ values.  

When considering the effect of stochastic perturbations on the control dynamics (i.e. $\omega_x=0, \omega_{\lambda}\neq0$) Figs.\ref{EffHLambda}a--c, the behavior of the 
efficiency remains qualitatively the same as in the purely deterministic case Fig.\ref{EffHopf}a. Regardless the intensity of control fluctuations, a maximum in 
efficiency is observed at an approximately same $\tau$ value. The efficiency associated with high and moderate control responses exhibit a similar 
dependency on $\tau$. In both cases, the maximum in
efficiency is higher as compared to the low response one. However, the highest efficiency peaks are slightly lowered when augmenting the intensity
of noise at the level of control. For all the control response cases, when increasing $\tau$ the corresponding efficiencies tend to converge to a same low value, independently of the intensity of noise in control.

Finally, in presence of strong fluctuations at the level of the dynamics of both the system and the control (i.e. $\omega_x\gg0, \omega_{\lambda}\gg0$), the efficiency rises towards a plateau for 
larger values of $\tau$. In this case it is observed that the presence of strong fluctuations enhances the control efficiency and to a greater extent in the cases of high and moderate control response.

\section{Summary and conclusions}

In this work we addressed the variance-based control of systems evolving towards criticality. As testbed models we considered the cusp and Hopf normal 
forms, displaying the prototypical phenomena of bistability and oscillations, respectively.

In relation to the core questions  \textbf{\textit{i)}}--\textbf{\textit{iii)}} exposed in Sec. I, our main results can be summarized as follows: 

\textbf{\textit{i)}} In absence of noise, effective control in both the cusp and the Hopf systems rises above a threshold value in the size of the observation time window. 

\textbf{\textit{ii)}} In the cusp-control system, changes in the observation time window provide a means to selecting among different control pathways, namely, hysteretic, branch-confined control, 
as well as an intermittent combination of both. The branch-confined regime becomes dominant when
considering a large observation window. This entails an increase in the complexity of the dynamics, in the form of aperiodic control-system trajectories. In the case of the Hopf-control 
system, the crossing of the bifurcation point could not be avoided, for the range of paramaters explored. Instead,
as the observation time window is enlarged, exploitation and recovery phases occur for longer periods. This gives rise to the alternace of time intervals where 
oscillations are suppressed, followed by amplified oscillations. 

\textit{\textbf{iii)}} In both systems, random fluctuations, at the level of the system variables, enhance the efficiency of control, and especially for large observation time windows. 
In the case of the cusp-control system, the intensity of noise induces a control pathway selection, similarly as in the deterministic case when varying the observation time window. From the 
deterministic to the strong fluctuations cases, the efficiency of the cusp-control system tends to increase for large values in the size of the observation window. In the Hopf-control system, an increase in the intensity of noise induces a reduction in the amplitude of 
oscillations, which entails an increase in the control efficiency.
Random fluctuations at the level of the control dynamics exert opposite influences on the control efficiency of the cusp and Hopf control systems. In the former 
it increases, while for the latter it tends to diminish the peak of efficiency. For both cusp and Hopf control systems, 
the combination of fluctuations in the control and in the system dynamics amplifies the efficiency in particular for large observation time windows.
Regarding the interplay between the efficiency, the degree of control response, the intensity of noise, and the observation time window, we observe: in both systems, the low control 
response case exhibits a low efficiency, independently of the intensity of noise and observation window. When increasing the variance of fluctuations in the dynamics of the cusp system,
 a moderate control 
response leads to a higher efficiency; the opposite situation occurs
 when considering fluctuations on the control dynamics. In the Hopf system, the moderate and high control response perform very similar, 
regardless the intensity of noise and size of the time window. In this situation, for both moderate and high response cases, the efficiency converges to 
the lower value associated with a low control response, as the observation time window is increased.

For the systems here addressed, it remains to be studied the influence of other parameters, such as the rate of change in the control system ($k_{\lambda}$),
the control reference parameter ($V_R$), as well as alternative approaches towards bifurcation, such as variation of the parameter $\mu$ 
in the cusp map --which would induce flickering patterns at the vicinity of the critical point. 

Potential applications of this 
approach to real world systems include ecosystem models, non-equilibrium mesoscopic physico-chemical systems or non-equilibrium models in economics, among others. Similarly, 
it is of interest to extend this work to address the control properties associated to different alternative
early warning indicators. Moreover, our analisis could be extended to address the control of coupled systems, forming networks whose nodes consist in dynamical systems subject to control.

Certainly, the structure of the evolution equations of real world models, involving a large number of parameters, lacks the degree of 
symmetry that confers to normal forms their characteristic simplicity. Moreover, not every dynamical system exhibiting a bifurcation belongs to a 
class of universality. It is thus important to address further the relation between the observation time scales leading to effective control and 
the geometrical properties, such as the curvature, of the system solution branches, while considering different early warning indicators. 
  
\section{Acknowledgements}
We acknowledge the Federal Ministry of Education and Research (BMBF) through the program 
``Spitzenforschung und Innovation in den Neuen L\"{a}nden'' (contract ``Potsdam Research Cluster for Georisk Analysis, 
Environmental Change and Sustainability'' D.1.1) and the
European Community's Seventh Framework Programme under Grant Agreement
No. 308497 (Project RAMSES) for financial support. A.G.C.R. acknowledges Christian Pape, Prof. Gregoire Nicolis and Doctors Elena Surovyatkina, Vasileos Basios and Flavio Pinto 
for fruitful discussions and for encouraging this work.

\bibliographystyle{unsrt}
%\bibliography{Bifcontrol}

\begin{thebibliography}{10}

\bibitem{NicolisI}
Nicolis G. and Prigogine I.
\newblock {\em Exploring Complexity: An Introduction}.
\newblock W. H. Freeman Press, 1989.

\bibitem{Politi}
R.~Badii and Politi A.
\newblock {\em Complexity: Hierarchical structures and scaling in physics}.
\newblock Cambridge Press, 1997.

\bibitem{Stanley}
H.~E. Stanley.
\newblock {\em Introduction to Phase Transitions and Critical Phenomena}.
\newblock Oxford Press, 1987.

\bibitem{NicolisII}
Nicolis G. and Prigogine I.
\newblock {\em Self-organization in Nonequilibrium Systems}.
\newblock John Wiley and Sons Inc., 1977.

\bibitem{Cross}
Cross M. and Greenside H.
\newblock {\em Pattern Formation and Dynamics in Nonequilibrium Systems}.
\newblock Cambridge Press, 2009.

\bibitem{SchefferIV}
Scheffer M., Carpenter S., Foley J.A., Folke C., and Walker B.
\newblock Catastrophic shifts in ecosystems.
\newblock {\em Nature}, 413:591--596, 2001.

\bibitem{Scheffer}
Scheffer M.
\newblock {\em Critical Transitions in Nature and Society}.
\newblock Princeton Press, 2009.

\bibitem{Castellano}
Castellano C., Fortunato S., and Loret V.
\newblock Statistical physics of social dynamics.
\newblock {\em Rev. Mod. Phys.}, 81:591--646, 2009.

\bibitem{Lenton}
Lenton~T. M., Held H., Kriegler E., Hall~J. W., Lucht W., Rahmstorf S., and
  Schellnhuber~H. J.
\newblock Tipping elements in the earth's climate system.
\newblock {\em Proc. Natl. Acad. Sci.}, 105:1786--1793, 2008.

\bibitem{Gardiner}
Gardiner C.
\newblock {\em Stochastic Methods: A Handbook for the Natural and Social
  Sciences}.
\newblock Springer-Verlag, 2009.

\bibitem{Guckenheimer}
Guckenheimer J. and Holmes P.
\newblock {\em Nonlinear Oscillations, Dynamical Systems, and Bifurcations of
  Vector Fields, Applied Mathematical Sciences 42}.
\newblock Springer-Verlag, 1983.

\bibitem{SchefferI}
Scheffer~M. et~al {et al}.
\newblock Early-warning signals for critical transitions.
\newblock {\em Nature}, 461:53--59, 2009.

\bibitem{SchefferII}
Scheffer~M. et~al {et al}.
\newblock Anticipating critical transitions.
\newblock {\em Science}, 338:344--348, 2012.

\bibitem{Carpenter}
Carpenter S. R. \& Brock~W. A.
\newblock Rising variance: a leading indicator of ecological transition.
\newblock {\em Ecology Letters}, 9:311--318, 2006.

\bibitem{Guttal}
Guttal V. and Jayaprakash C.
\newblock Changing skewness: an early warning signal of regime shifts in
  ecosystems.
\newblock {\em Ecology Letters}, 11:450--460, 2008.

\bibitem{Biggs}
Biggs R., Carpenter~S. R., and Brock~W. A.
\newblock Turning back from the brink: Detecting an impending regime shift in
  time to avert it.
\newblock {\em Proc. Nat. Acad. Sci. USA}, 106:826--831, 2009.

\bibitem{Livina}
Livina V.N. and Lenton~T. M.
\newblock A modified method for detecting incipient bifurcations in a dynamical
  system.
\newblock {\em Geophys. Res. Lett}, 34:L03712, 2007.

\bibitem{Dakos}
Dakos~V. et~al {et al}.
\newblock Methods for detecting early warnings of critical transitions in time
  series illustrated using simulated ecological data.
\newblock {\em PLoS ONE}, 7:e41010, 2012.

\bibitem{Ray}
Hilborn R.
\newblock Reinterpreting the state of fisheries and their management.
\newblock {\em Ecosystems}, 10:1362--1369, 2007.

\bibitem{SchefferIII}
Scheffer M., Hosper~S. H., Meijer~M. L., and B.~Moss.
\newblock Alternative equilibria in shallow lakes.
\newblock {\em Trends Ecol. Evol.}, 8:275--279, 1993.

\bibitem{NicolisIII}
Nicolis G. and Nicolis C.
\newblock Dynamics of switching in nonlinear kinetics.
\newblock {\em J. Phys.: Condens. Matter}, 19:065131, 2007.

\bibitem{NicolisIV}
Nicolis G.
\newblock {\em Introduction to Nonlinear Science}.
\newblock Cambridge Press, 1999.

\bibitem{DakosI}
V.~Dakos, van Nes~E.H., D'Odorico P., and Scheffer M.
\newblock Tipping elements in the earth's climate system.
\newblock {\em Ecology}, 9:264--271, 2012.

\bibitem{Hastings}
Boettiger C. and Hastings A.
\newblock Quantifying limits to detection of early warning for critical
  transitions.
\newblock {\em Interface}, 9:2527--2539, 2012.

\bibitem{McEwen}
McEwen J.S., Garc{\'i}a Cant{\'u}~Ros A., Gaspard P., Visart de~Bocarme~T., and
  Kruse N.
\newblock Non-equilibrium surface pattern formation during catalytic reactions
  with nanoscale resolution: Investigations of the electric field influence.
\newblock {\em Catalysis Today}, 154:75--78, 2010.

\bibitem{Epstein}
Epstein I.R. and Pojman J.A.
\newblock {\em An introduction to nonlinear chemical dynamics: oscillations,
  waves, patterns, and chaos}.
\newblock Oxford Press, 1998.

\bibitem{Goldbeter}
Goldbeter A.
\newblock {\em Biochemical Oscillations and Cellular Rhythms: The Molecular
  Bases of Periodic and Chaotic Behaviour}.
\newblock Cambridge Press, 1997.

\bibitem{Murray}
Murray G.D.
\newblock {\em Mathematical Biology: I. An Introduction}.
\newblock Springer-Verlag, 2000.

\bibitem{Vallis}
Vallis G.K.
\newblock Conceptual models of el ni{\~n}o and the southern oscillation.
\newblock {\em J. Geophys. Res.}, 93:13979--13991, 1988.

\bibitem{Hallegatte}
Hallegatte S., Ghil M., Dumas P., and Hourcade J.C.
\newblock Business cycles, bifurcations and chaos in a neo-classical model with
  investment dynamics.
\newblock {\em Journal of Economic Behavior and Organization}, 67:57--77, 2008.

\end{thebibliography}

\end{document}